\documentclass{PoS}
\usepackage{multirow}

\newcommand{\gev}{\,\mbox{GeV}}
\newcommand{\ams}{\alpha_{\overline{MS}}}
\newcommand{\amv}{\alpha_V}

\title{Matching the Bare and $\overline{MS}$ Charm Quark Masses Using
  Weak Coupling Simulations}

\ShortTitle{Matching $m_0$ and $m_{\overline{MS}}$}

\author{\speaker{I.~F.~Allison}$^a$~\thanks{Email:iana@triumf.ca},
  K.~Y.~Wong$^b$, C.~T.~H.~Davies$^b$, C.~McNeile$^b$,
  H.~D.~Trottier$^c$, E.~Dalgic$^c$, J.~Wu$^a$, E.~Follana$^d$,
  R.~R.~Horgan$^e$,
  G.~P.~Lepage$^f$, J.~Shigemitsu$^g$,\\
  for the HPQCD collaboration\\\\
  \llap{$^a$} TRIUMF, 4004 Wesbrook Mall, Vancouver, BC, V6T 2A3, Canada \\
  \llap{$^b$} Department of Physics and Astronomy, University of
  Glasgow,
  Glasgow, G12 8QQ,  UK\\
  \llap{$^c$} Department of Physics, Simon Fraser University, Vancouver, BC, V5A 1S6, Canada\\
  \llap{$^d$} Department of Theoretical Physics, University of Zaragoza, E-50009, Zaragoza, Spain, \\
  \llap{$^e$} DAMTP, CMS, University of Cambridge, Cambridge, CB3 0WQ, UK \\
  \llap{$^f$} Laboratory for Elementary-Particle Physics, Cornell University, Ithaca, NY, 14853, USA \\
  \llap{$^g$} Physics Department, The Ohio State University, Columbis,
  Ohio, 43210, USA}
\abstract{We provide a new determination of the charm quark mass using
  the Highly Improved Staggered Quark (HISQ) action, finding
  $m_c^{\overline{MS}}(3\gev) = 0.983(23)\gev$. Our determination
  makes extensive use of second order lattice perturbation theory in
  matching the bare lattice mass to the $\overline{MS}$ scheme. This
  matching utilises both traditional diagrammatic perturbation theory
  and weak coupling simulations. The second of these techniques allows
  us to extract perturbative coefficients from Monte-Carlo simulations
  and the process of doing this is laid out in some detail here.}

\FullConference{The XXVI International Symposium on Lattice Field Theory\\
  July 14-19 2008\\
  Williamsburg, Virginia, USA}

\begin{document}
\section{Introduction}\label{sec:introduction}
The quark masses are important both as fundamental parameters and,
more pragmatically, as inputs to experimental determinations of CKM
matrix elements~\cite{Ahrens:2008qu}. The charm quark is particularly
important because of the large flavour physics program, but it has
been somewhat neglected on the lattice due to the difficulty in
simulating it accurately. In these proceedings, we make use of the
highly-improved staggered quark (HISQ) action~\cite{Follana:2006rc} to
extract a value for the charm quark mass from dynamical lattice
QCD. Empirically, the HISQ action is known to reduce the
$\mathcal{O}(\alpha_sa^2)$ errors which remain in the AsqTad action,
and which are thought to be taste-changing errors, making precision
charm physics possible~\cite{Follana:2006rc}. It does this by
repeating the AsqTad link smearing, which further suppresses taste
changing interactions, and also by correcting the dispersion relation
through adjustment of the N\"aik term coefficient. In this work, we
aim to use these features to calculate the charm quark mass from
$\eta_c$ correlators on the lattice.

In quoting a determination of a quantity such as the charm quark mass,
it is customary to convert to the $\overline{MS}$ renormalization
scheme and to use a standard scale (e.g. $3\gev$). To do this directly
from a determination of the bare lattice QCD mass requires lattice
perturbation theory. The trend towards increasingly complicated
actions, such as HISQ, has made most calculations of this type a major
computational undertaking. One tool developed in response to this is
the use of weak coupling or \emph{high}-$\beta$
simulations~\cite{Dimm:1994fy}. At sufficiently large values of the
coupling $\beta$ (equivalently, sufficiently small lattice spacings) a
lattice simulation will have a small physical volume and a very large
cutoff($\approx\pi/a$). These are precisely the conditions required to
probe the perturbative regime of QCD and when perturbation theory is
done in this way (by Monte-Carlo) all orders are automatically
included. When it is used in combination with a technique like
constrained curve fitting, the high-$\beta$ technique can allow
diagrammatic results to be extended to the next order at the cost of
running some extra simulations, provided the quality of the
high-$\beta$ results is sufficient. This approach has been
successfully demonstrated in~\cite{Wong:2005jx}. Of course, there are
some complications, mostly related to the very small volumes of the
simulations, but the two most significant problems: the existence of
zero modes and $\mathbb{Z}_3$ tunneling, are known to be effectively
resolved by the use of (color) twisted boundary
conditions\cite{Luscher:1985wf}, see e.g.~\cite{Trottier:2001vj}. In
this work, we will use high-$\beta$ simulations with twisted boundary
conditions on all of the spatial dimensions to do part of the second
order matching.

The recently published determination of~\cite{Allison:2008xk} also
used HISQ quarks and a mixture of continuum and lattice techniques to
calculate the charm quark mass, finding $m_{\overline{MS}}(3\gev) =
0.986(10)\gev$. The calculation we will present uses a completely
different method, extracting $m_c$ from $\eta_c$ correlators before
manually matching to the $\overline{MS}$ scheme. Together we view
these independent calculations as giving important cross checks of one
another.

\section{Matching to the $\overline{MS}$
  Scheme}\label{sec:details-high-beta}
The lattice charm quark mass $am_c$ can be matched to the $\overline{MS}$
scheme mass $m_{\overline{MS}}$ using the on-shell mass $M$ as an
intermediate stage.
\begin{eqnarray}
  \label{eq:2}
  m_{\overline{MS}}(\mu) &=& M\left[
    1 + \left(B_{11}l + B_{10}\right)\ams(\mu) +
    \left(B_{22}l^2 + B_{21}l + B_{20}\right)\ams^2(\mu)
  \right] + \mathcal{O}(\ams^3) \nonumber\\
  M &=& am_c\left[
    1 + 
    \left(A_{11}L + A_{10}\right)\alpha_L +
    \left(A_{22}L^2 + A_{21}L + A_{20}\right)\alpha_L^2\right] +
  \mathcal{O}(\alpha_L^3),
\end{eqnarray} 
where $L=\log{am_c}$ and $l = \log{\mu}/{M}$. The relation between
$m_{\overline{MS}}$ and $M$ (the $B$ coefficients) is given to third
order in~\cite{Chetyrkin:2000uy} so only the $A$ coefficients are
unknown. The connections between $\alpha_L\to\alpha_V$, and
$\ams\to\alpha_V$ are given in~\cite{Trottier:2003bw}
and~\cite{Schroder:1998vy} respectively.  Writing $m_{\overline{MS}}$
in terms of $am_c$ to second order in $\alpha_V$ and demanding that
the unphysical dependence on $L=\log{am_c}$ vanishes gives conditions
on the coefficients $A_{11}, A_{22}$ and $A_{21}$ which result in the
form
\begin{eqnarray}
  \label{eq:10}
  m_{\overline{MS}}(\mu) 
  &=&  am_c \left(
    1 + 
    \left(Z_{11}l_{a\mu } + Z_{10}\right)\amv(aq^*) + 
    \left(Z_{22}l^2_{a\mu } + Z_{21}l_{a\mu } + Z_{20}\right)\amv^2(aq^*)
  \right) + \mathcal{O}(\amv^3) \nonumber\\
  &=& am_c
  + c_1(m_qa)\alpha_V(aq^*) + 
  \left(c_{2,q} + c_{2,g}\right)\alpha_V^2(aq^*)
  + \mathcal{O}(\alpha_V^3),
\end{eqnarray}
with $l_{a\mu}=\log{a\mu}$ and
\begin{eqnarray}
  \label{eq:16}
  Z_{11} &=& -\frac{2}{\pi}, \qquad 
  Z_{10} = A_{10}-\frac{4}{3\pi}, \qquad 
  Z_{22} = \frac{15}{2\pi^2} - 
  \frac{n_f}{3\pi^2}, \nonumber\\
  Z_{21} &=& \left(\frac{2\log{aq}}{3\pi^2} - 
    \frac{5}{18\pi^2}\right)n_f - 
  \frac{11\log{aq}}{\pi^2} - 
  \frac{7}{12\pi^2} - 
  \frac{2A_{10}}{\pi}, \nonumber\\
  Z_{20} &=& \left( 
    \frac{\log{\pi/aq}}{3\pi}A_{10} + 
    \frac{4\log{aq}}{9\pi^2} +
    \frac{53}{432\pi^2} + 
    \frac{1}{18}\right)n_f + 
  A_{20} \nonumber\\
  &&+ 
  \left(
    \frac{2}{3\pi} - 
    \frac{11\log{\pi/aq}}{2\pi} - 
    v_{1,0}
  \right)A_{10} + 
  \frac{\zeta_3}{6\pi^2} - 
  \frac{2 + \log{2}}{9} -
  \frac{22\log{aq}}{3\pi^2} - 
  \frac{257}{32\pi^2}.
\end{eqnarray}
The splitting of the fermionic and gluonic portions of $c_2$ in the
second line of equation~(\ref{eq:10}) is motivated by there being only
$4$ fermionic diagrams for the second order mass
renormalization. These diagrams have been evaluated using diagrammatic
perturbation theory (see~\cite{Dalgic:2007zz} for an outline of this
calculation, final results are in preparation). The remaining diagrams
which contribute to $c_{2,g}$ represent a much larger undertaking and
are our motivation for the use of the high-$\beta$ technique. One
advantage of the split is that we only require \emph{quenched} results
for $c_{2,g}$. The remaining unknown coefficients $A_{10}$ and
$A_{20}=A_{20,g}+A_{20,f}$, can then be expressed in the following way
\begin{eqnarray}
  \label{eq:17}
  A_{10} &=& \frac{c_1}{am_c} + 
  \frac{2}{\pi}L,\\
  A_{20,g} & = & 
  \left(\frac{L^2}{3\pi^2} -
    \frac{1 + 4\log{\pi}}{6\pi^2}L-
    \frac{c_1\log{\pi/aq}}{3\pi am_c}\right)n_f - 
  \frac{7}{2\pi^2}L^2 + \nonumber \\
  && + \left(
    \frac{2c_1}{\pi}+\frac{79+132\log{\pi}+24\pi v_{1,0}}{12\pi^2}
  \right)L 
  + \frac{2\pi c_{2,g} + 11 c_1 \log{\pi/aq} + 2\pi c_1 v_{1,0}}{2am_c}.
\end{eqnarray}
$A_{20,f}$ comes from the diagrammatic analysis
of~\cite{Dalgic:2007zz}. Together with equation~(\ref{eq:16}) these
coefficients allows us to evaluate equation~(\ref{eq:10}) and extract
a physical values of $m_{\overline{MS}}$.

\section{Results}
\subsection{The Lattice Charm Quark Mass}
The lattice bare charm quark mass was tuned by adjusting it until the
$\eta_c$ mass agreed with experiment on four ensembles of the MILC
collaboration's configurations~\cite{Bernard:2001av}. In this tuning
the scale was set using MILC values of $r_1/a$ and the value $r_1 =
0.321(5)\,\mbox{fm}$~\cite{Gray:2005ur} . The bare mass was adjusted
for any mistuning (a very small effect in all cases) and then
converted to the bare tree level mass\footnote{The tree level mass is
  related to the bare mass via $m_{tree} = m_0\left(1 -
    \frac{3}{80}m_0^4 + \frac{23}{2240}m_0^6 +
    \frac{1783}{573600}m_0^8 - \frac{76943}{23654400}m_0^{10} +
    \mathcal{O}(m_0^{12})\right)$ for HISQ. This relation can be
  determined from the free field HISQ action.} which is the quantity
related to the $\overline{MS}$ mass in equation~(\ref{eq:10}). The
$m_{c,tree}$ values are given in Table~\ref{tab:mcbare}.
\begin{table}[ht]
  \centering
  \begin{tabular}{ccc|c|c|c|c|c}
    \hline\hline
    Size & $u_0am_l$ & $u_0am_s$ & $am_c$ & $am_{\eta_c}$ & 
    $m_{c,tree}/\gev$ & $1+\epsilon$ & $r_1/a$ \\
    \hline
    \multirow{2}{*}{$16^3\times 48$} & $0.0194$ & $0.0484$ & 
    $0.85$ & $2.26964(17)$ & $1.100(2)(2)$ & $0.66$ &$2.129(11)$ \\
    & $0.0097$ & $0.0484$ &  $0.85$ & $2.27031(16)$ & $1.098(0)(2)$ & 
    $0.66$ & $2.133(11)$ \\
    \hline
    \multirow{2}{*}{$20^3\times 64$} & $0.02$ & $0.05$ & 
    $0.648$ & $1.84153(17)$ & $1.040(5)(1)$ & $0.79$ & $2.650(8)$ \\
    & $0.01$ & $0.05$ & $0.66$ & $1.87142(12)$ & $1.041(6)(2)$ &
    $0.79$ & $2.610(12)$ \\
    \hline
    $24^3\times 64$ & $0.005$ & $0.05$ & $0.65$ & $1.84949(11)$ & 
    $1.039(3)(2)$ & $0.79$& $2.632(13)$ \\
    \hline
    \multirow{2}{*}{$28^3\times 96$} & $0.0124$ & $0.031$ &
    $0.427$ & $1.30731(11)$ & $0.9718(5)(11)$ & $0.885$ & $3.711(13)$ \\
    & $0.0062$ & $0.031$ & $0.43$ & $1.31693(12)$ & $0.9715(5)(11)$ & 
    $0.885$ & $3.684(12)$ \\
    \hline
    $48^3\times 144$ & $0.0036$ & $0.018$ & $0.28$ & $0.91555(8)$ & $0.9129(25)(9)$ &
    $0.949$ & $5.277(16)$ \\
    \hline\hline
  \end{tabular}
  \caption{Simulation parameters for extracting the lattice charm quark mass
    The value of the correction of the n\"aik term,
    $\epsilon$ used here was determined non-perturbatively forcing 
    the ``speed of light'' to be 1. This differs, but not significantly, 
    from the series definition of $\epsilon$ used in the perturbative 
    portion of our calculations. The value of $r_1$, used to set the scale,
    was taken to be $r_1 = 0.321(5)\,\mbox{fm}$}
  \label{tab:mcbare}
\end{table}

\subsection{High-$\beta$ Perturbative Results}\label{sec:perturbation-theory}
\begin{table}[ht]
  \centering
  \begin{tabular}{ccc}
    \hline\hline
    & HISQ & ASQTAD \\
    \hline
    $L^3\times T$ & $6^3\times 16$, $8^3\times 20$, 
    $10^3\times$, $12^3\times 20$ & 
    $6^3\times 16$, $8^3\times 20$, 
    $10^3\times$, $12^3\times 20$ \\
    $\beta$ & $15, 16, 20, 24, 32, 46, 62, 70, 92$ &
    $15, 16, 20, 24, 32, 46, 62, 70, 92$ \\
    $m_0$ & $0.30, 0.43, 0.50, 0.66, 0.85$ & 
    $0.30, 0.40, 0.50, 0.60, 0.70$ \\
    \hline
  \end{tabular}
  \caption{Parameters for the high-$\beta$ simulations.}
  \label{tab:highbetaparam}
\end{table}
We performed high-$\beta$ simulations for valence HISQ and AsqTad
quarks at the parameters given in table~\ref{tab:highbetaparam}.
To extract $c_1$ and $c_{2,g}$ we started from the on shell mass
$M(L,\beta)$ determined by simulating quark propagators in a
Coulomb+Axial gauge\footnote{This is not the traditional maximal-tree
  gauge but is modified to take account of the twisted boundary
  conditions.}, and then fitting to the form
\begin{equation}
  \label{eq:14}
  aM_{pole} = E_1 + 
  c_1\alpha_V(aq^*) + 
  c_{2,g}\alpha_V^2(aq^*) + 
  \cdots.
\end{equation}
Because high-$\beta$ results are essentially perturbations around the
free field, we used constrained curve fitting with the first term set
to the free field energy of the HISQ action, allowing us to evaluate
finite volume values for $c_1$ and $c_{2,g}$. The values of
$\alpha_V(aq^*)$ were evaluated for each simulation by measuring the
plaquette and using the three loop expansion of $\log_{W_{1\times 1}}$
given in~\cite{Wong:2005jx} to extract $\alpha_V(q^*_{plaq})$ which
was then evolved to the $q^*$ relevant to our simulations.

\subsection{Comparison of HISQ $c_1$ with Diagrammatic Perturbation Theory}
\begin{figure}[htb]
  \centering
  \begin{minipage}[c]{0.42\textwidth}
    \centering
    \begin{displaymath}
      \label{eq:15}
      c_1(L) = c_1(L=\infty) + 
      \frac{X_{c,1}}{L} + 
      \frac{X_{c,2}}{L^2} + 
      \cdots
    \end{displaymath}\\
    \vspace{2mm}
    \begin{tabular}{c|cc}
      \hline\hline
      mass & $c_1^{MC}$ & $c_1^{PT}$ \\
      \hline
      $0.30$ & $0.4477(17)$ &  $0.4505(7)$ \\
      $0.43$ & $0.4932(12)$ & $0.4921(7)$ \\
      $0.66$ & $0.5979(15)$ & $0.5978(7)$ \\
      $0.85$ & $0.6634(15)$ & $0.6693(7)$ \\
      \hline
    \end{tabular}
  \end{minipage}
  \begin{minipage}[c]{0.56\textwidth}
    \centering
    \includegraphics[width=0.85\linewidth]{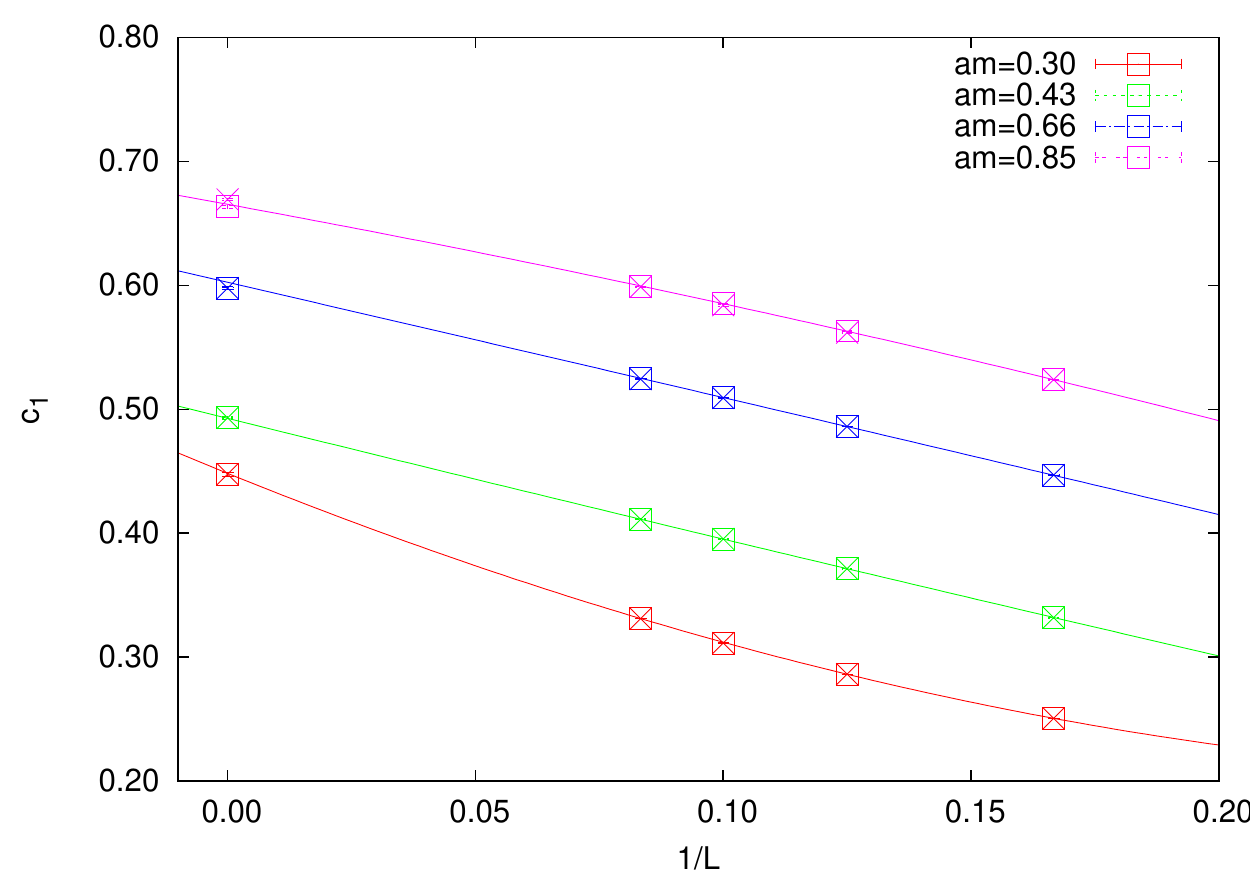}
  \end{minipage}
  \caption{$c_1$ infinite volume extrapolations. Boxes are
    high-$\beta$ results, crosses are diagrammatic PT results. The
    errors come from bootstrapping the entire analysis with 1000
    bootstrap re-samples.}
  \label{fig:c1infvol}
\end{figure}

Fits to equation~(\ref{eq:14}) were performed including terms up to
$\mathcal{O}(\alpha_V^4)$ with priors of $0\pm5$ for all of the
$c_i$. We used the resulting $c_1$ values as a check on our method by
comparing to the corresponding finite volume diagrammatic perturbation
theory values as shown in figure~\ref{fig:c1infvol}. Our results were
then extrapolated to the infinite volume limit, where we included
terms up to fourth order in the fits with priors of $0\pm3$ for all
parameters. These results were also compared to diagrammatic
perturbation theory results and are again shown in
figure~\ref{fig:c1infvol}.

\subsection{Comparison of AsqTad $c_{2,g}$ with Diagrammatic Perturbation Theory}
\begin{figure}[htb]
  \centering
  \begin{tabular}{c|cc}
    \hline\hline
    mass & $c^{MC}_{2,g}$ & $c^{PT}_{2,g}$ \\
    \hline
    $0.30$ & $1.182(94)$ & $1.00(2)$ \\
    $0.40$ & $1.350(99)$ & $1.22(3)$ \\
    $0.60$ & $1.72(11)$ & $1.65(3)$ \\
    $0.70$ & $2.02(13)$ & $2.12(4)$ \\
    \hline
  \end{tabular}
  \begin{minipage}[c]{0.38\textwidth}
    \centering
  \end{minipage}
  \begin{minipage}[c]{0.58\textwidth}
    \centering
    \includegraphics[width=0.85\linewidth]{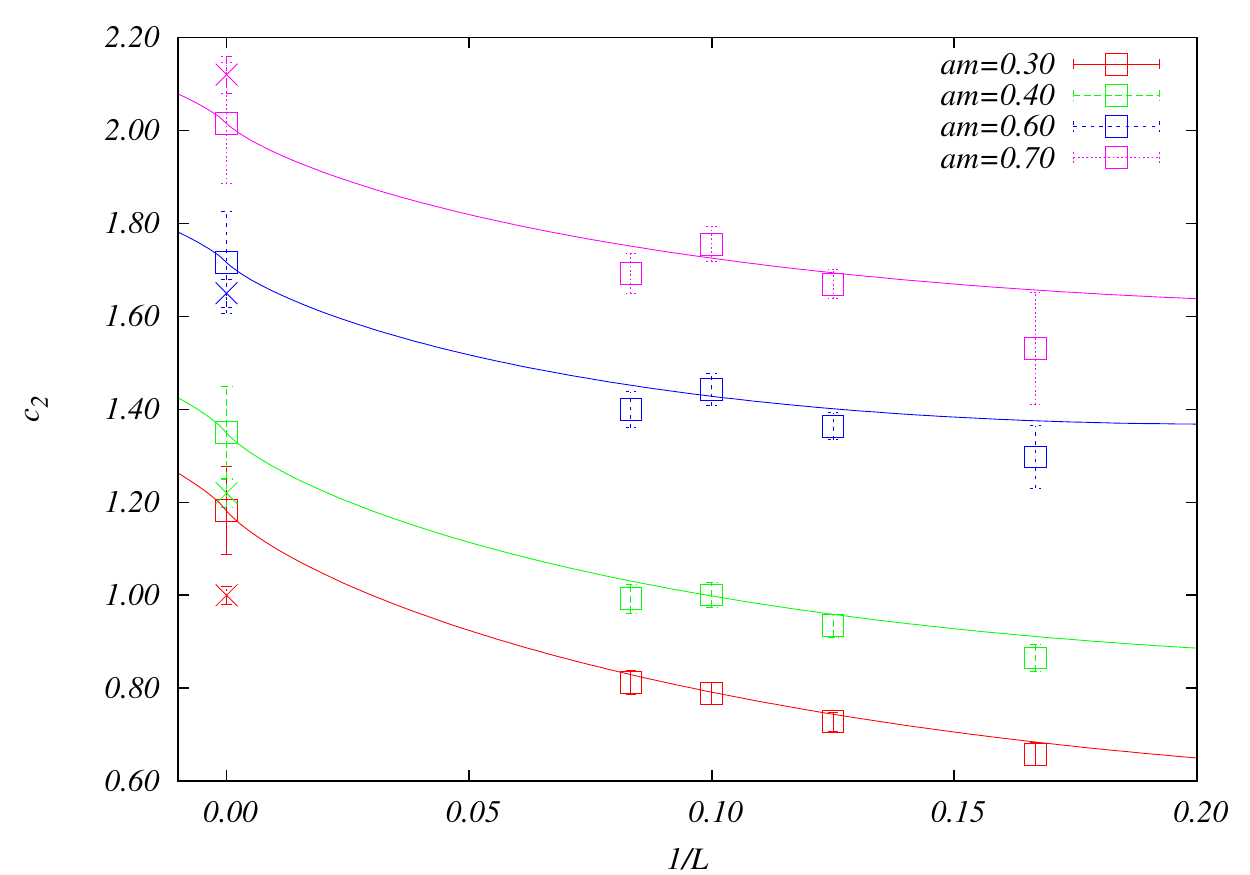}
  \end{minipage}
  \caption{AsqTad $c_{2,g}$ infinite volume extrapolations. The boxes are
    our high-$\beta$ results and the corresponding infinite volume
    extrapolation. The crosses are diagrammatic perturbation theory
    results. For this check, the analysis was \emph{not} bootstrapped,
    errors are fitting/statistical only.}
  \label{tab:c2infvol}    
\end{figure}

For AsqTad valence quarks, $c_1$ and $c_{2,g}$ have already been
calculated using diagrammatic perturbation theory
in~\cite{Mason:2005bj}. We used finite volume values of $c_1$ as
priors to aid our extraction of $c_{2,g}$ and then extrapolated to the
infinite volume limit via\footnote{The justification for this form for
  the extrapolation comes from~\cite{Trottier:2001vj}.}
\begin{equation}
  \label{eq:8}
  c_{2,g}(L)=c_{2,g}(L=\infty) + \frac{1}{L}(X_{c_2,1} + Y_{c_2,1}\log{L^2}) + 
  \frac{1}{L^2}(X_{c_2,2}+Y_{c_2,2}\log{L^2}) + \cdots
\end{equation}
where $Y_{c_2,1} = \frac{11}{4\pi}X_{c_1,1}$ and $X_{c_1,1}$ is the
same quantity which appears in extrapolation of $c_1$ and which we
were able to use as a further constraint in our fits. The results of
these fits are given in figure~\ref{tab:c2infvol} and we interpret
them as lending weight to our $c_{2,g}$ calculation for HISQ.

\section{HISQ $c_{2,g}$ From High-$\beta$}
\begin{figure}[htb]
  \centering
  \begin{minipage}[c]{0.38\textwidth}
    \centering
    \begin{tabular}{c|c}
      \hline\hline
      mass & $c_{2,g}$ \\
      \hline
      $0.30$ & $0.327(34)$ \\
      $0.43$ & $0.515(38)$ \\
      $0.66$ & $0.130(56)$ \\
      $0.85$ & $-0.438(63)$ \\
      \hline
    \end{tabular}
  \end{minipage}
  \begin{minipage}[c]{0.58\textwidth}
    \centering
    \includegraphics[width=0.85\linewidth]{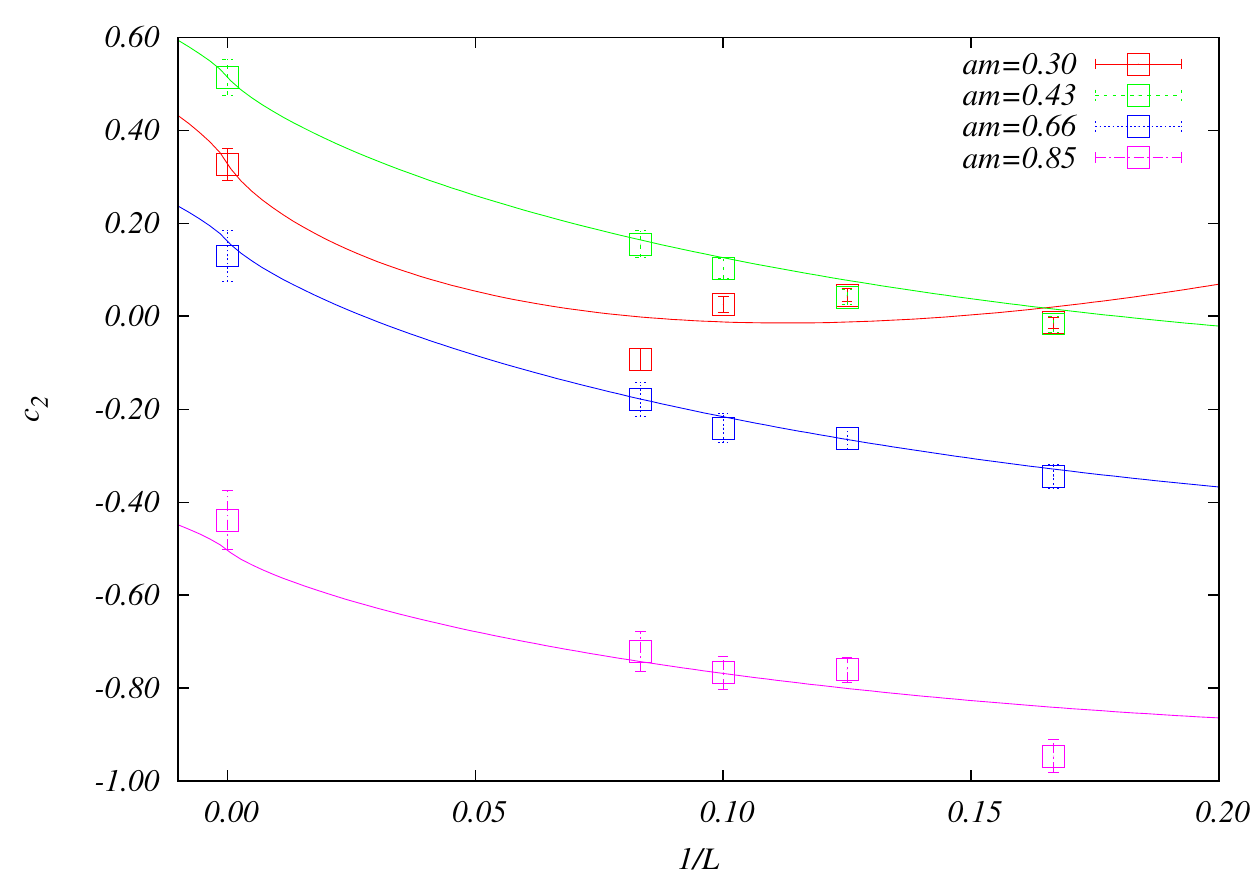}
  \end{minipage}
  \caption{$c_{2,g}$ Infinite volume extrapolation for HISQ. The curved
    lines in the plot are fits to equation 3.3. The errors come from
    bootstrapping the analysis with 1000 bootstrap re-samples.}
  \label{fig:c2infvolHISQ}    
\end{figure}

The new result which we present here is a determination of the gluonic
part of the $c_2$ coefficient for HISQ, for which there are no
corresponding diagrammatic perturbation theory results. Again we
constrained the finite volume $c_1$ coefficients with diagrammatic
results and $X_{c_1,1}$ from our $c_1$ fits. The final results are
shown in figure~\ref{fig:c2infvolHISQ}. The results are encouraging
with the possible exception of the result for $am=0.30$ which may be
affected by finite volume corrections. We are currently investigating
this possibility by running at larger volumes.

\section{Conclusions}\label{sec:conclusions}
\begin{figure}[htb]
  \centering
  \begin{minipage}[c]{0.35\textwidth}
    \centering
    \begin{tabular}{ccc}
      \hline\hline
      & mass & $m_c^{\overline{MS}}(3\gev)$ \\
      \hline
      v.coarse &$0.28$ & $0.9729(53)$ \\
      coarse   &$0.43$ & $0.9777(12)$ \\
      fine     &$0.66$ & $0.9745(59)$ \\
      s.fine   &$0.85$ & $0.9774(18)$ \\
      \hline
      continuum & -- & $0.9830(64)$ \\
      \hline
    \end{tabular}

  \end{minipage}
  \begin{minipage}[c]{0.62\textwidth}
    \centering
    \includegraphics[width=0.8\linewidth]{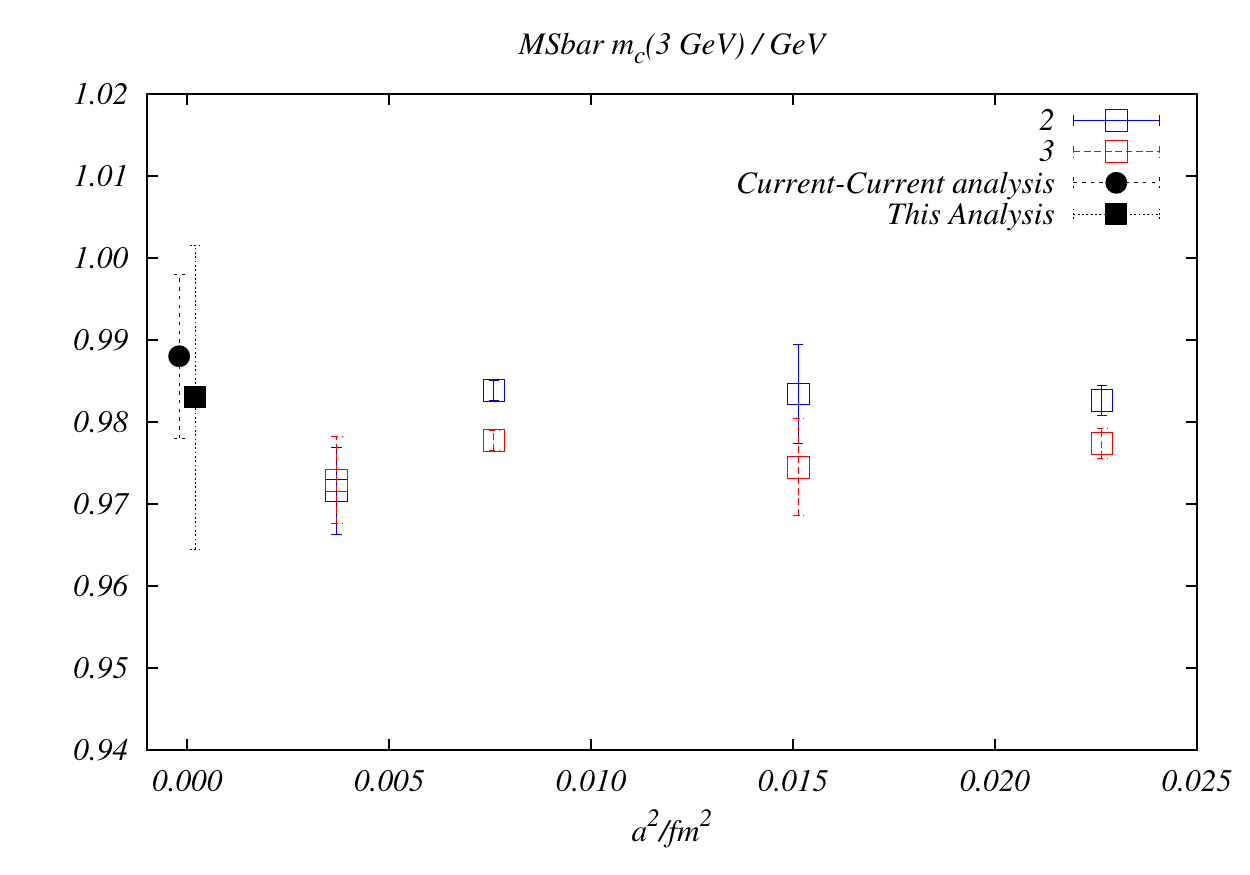}
  \end{minipage}
  \caption{Blue points come from full second order analysis, red
    points come from fits which include a parametrization of the third
    order terms. The black circle comes from the analysis
    of~\cite{Allison:2008xk}.}
  \label{fig:charmqm}
\end{figure}

The high-$\beta$ method has allowed us to extract a second order
perturbative coefficient which would otherwise have been a \emph{very}
expensive calculation in diagrammatic perturbation theory. Together
with calculations of $c_1$, $c_{2,q}$ and equation~(\ref{eq:10}) this
allows us to provide a two loop determination of the charm quark mass
using HISQ valence quarks. The final result for each lattice spacing
we used is given in figure~\ref{fig:charmqm}. The quoted continuum
value comes from fitting all lattice spacings simultaneously while
demanding a single common charm quark mass, allowing for higher order
perturbative and discretization errors.

Beyond statistical/fitting errors, the other important sources of
error are the orders excluded in the perturbative matching and the
overall scale determination. We estimated the error from the
perturbative matching by repeating our analysis but including the
third order perturbative coefficients with $A_{30}$ floated as a very
wide prior ($0\pm45$). The resulting determinations are also shown in
figure~\ref{fig:charmqm} and suggest that estimating missing order
terms by twice a typical value of $\alpha_V^3\approx0.22^3$ used in
the matching is conservative. The error from setting the overall scale
(which we do via $r_1$) was estimated as $0.5\%$ from an overall error
of $1.5\%$ on $r_1$ because the $r_1$ error affects only the binding
energy for the $\eta_c$ (see~\cite{Allison:2008xk}). Including these
sources of error, our preliminary result is
\begin{equation}
  \label{eq:18}
  m_c^{\overline{MS}}(\mu=3\gev) = 0.9830(64)(49)(213)\gev 
  \qquad\mbox{(stat./fitting)(scale)(higher orders)}.
\end{equation}
A more systematic determination, including better estimates of the
effect of missing higher order matching and chiral effects is ongoing
and will appear subsequently. At present though, our preliminary
result is in very good agreement with the determination
of~\cite{Allison:2008xk} though with a slightly larger error. We
interpret this result as a striking demonstration of the capabilities
of modern lattice simulations using highly improved actions such as
HISQ to give precise and physically relevant results needed by the
rest of the particle physics community.

\end{document}